\documentclass[12pt]{article}   
\usepackage{amsmath,graphicx}
\usepackage{amssymb, amsxtra} 
\usepackage{blkarray}
\usepackage{mathtools}
\usepackage{url}

\usepackage{yfonts}
\def\therefore{
\leavevmode
\lower0.1ex\hbox{$\bullet$}
\kern-0.2em\raise0.9ex\hbox{$\bullet$}
\kern-0.2em\lower0.2ex\hbox{$\bullet$}
\thinspace}

\title{On CP-violation and quark masses: reducing the number of parameters}
\author{A. Kleppe\footnote{sactacmk@gmail.com}\\ SACT, Oslo}
\date{}
\begin{document}
\maketitle

\begin{abstract}
A physically viable ansatz for quark mass matrices must satisfy certain constraints. In this article we study a concrete example, by looking at some generic matrices with a nearly democratic texture, and the implications of the constraint imposed by CP-violation, specifically
the Jarlskog invariant. We find that the number of mass parameters is reduced from six to five, implying that the six mass eigenvalues of the up-quarks and the down-quarks are interdependent, which in our approach is explicitly demonstrated.
\end{abstract}
\section{Introduction} 
A mass matrix ansatz is a suggestion of what form the quark mass matrices may have in the weak (flavour) basis. The hope is to find mass matrices that could shed some light on the enigmatic mass spectra. In this article, we study the constraints imposed by CP-violation on the quark mass matrices, using the mathematical tool provided by the Jarlskog invariant\cite{comm}. 

The usual ``mathematical reason'' given for CP-violation, is that the 3 $\times$ 3 weak mixing matrix $V_{CKM}$\cite{CKM} has a phase that cannot be rotated away, but in the 1980s, Cecilia Jarlskog discovered that a signature of CP-violation is that (determinant of) the commutator of the mass matrices of the up- and down-sectors is nonzero, or $det[M_u,M_d] \neq 0$, where $M_u$ and $M_d$ are the mass matrices of the up-sector and down-sector, respectively. From this she derived a
direct measure of weak CP-violation, namely the Jarlskog invariant
\begin{equation}
J_{CP}=-i\hspace{1mm} det[M_u,M_d]/2 P_u P_d 
\end{equation}
where $P_u=(m_u-m_c)(m_c-m_t)(m_t-m_u)$, $P_d=(m_d-m_s)(m_s-m_b)(m_b-m_d)$, $M_u,M_d$ are non-commuting hermitian matrices\footnote{The rigorous expression is $J_{CP}=-i\hspace{1mm}det[M_uM_u^{\dagger},M_dM_d^{\dagger}]/2(m_u^2-m_c^2)...(m_d^2-m_s^2)...$, to ensure hermiticity, but the matrices we are going to use are hermitian, so we use the simpler version.}, and $m_j$ are the mass eigenvalues.

Technically speaking, the weak CP-violation is related to the complex elements in the weak mixing matrix, and the connection between the weak mixing matrix and the Jarlskog invariant can be expressed as
\begin{equation}
J_{CP}=Im(V_{ij} V_{kl} V^{*}_{kj} V^{*} _{il})  
\end{equation}
where $V_{ij}$, are the matrix elements of the mixing matrix, and $i,j,k,l=1,2,3$.
\newline
To calculate the Jarlskog invariant $J_{CP}$, we can use the Wolfenstein parametrization\cite{wolf} of the 
weak mixing matrix,
\begin{equation}\label{Wol}V_{Wolf}=
\begin{pmatrix}
1-\lambda^2/2             &  \lambda     &  A\lambda^3(\rho-i\eta)\\
-\lambda                 & 1-\lambda^2/2 &  A\lambda^2 \\
A\lambda^3(1-\rho-i\eta)  & -A\lambda^2  & 1\\
\end{pmatrix}
\end{equation}
where $\lambda=0.2245$, $A=0.836$, $\rho=0.122$, $\eta=0.355$.
Inserting the mixing matrix elements for these values in the expression $J_{CP}=Im(V_{ij} V_{kl} V^{*}_{kj} V^{*} _{il})  $, we get
$J_{CP}= 3.096\times 10^{-5}$, in agreement with the value given by the Particle Data group\cite{PDG}, $J_{CP}= (3.18\pm 0.15)\times 10^{-5}$.

\section{Mass matrices}
The Jarlskog invariant implies that in order to be meaningful, an ansatz for the quark mass matrices must provide an explicit matrix ansatz for both quark charge sectors. Only then can we ensure that their commutator satifies the constraint imposed by $J_{CP}$, and the very first step is obviously to make sure that the commutator has a non-vanishing determinant. For the sake of concreteness, we here study the implications of the Jarlskog invariant for some rather generic matrices. 

In an earlier article \cite{Koi}, we studied matrices with a certain, nearly democratic structure, with the purpose of investigating the relations between the mass matrices for the two quark sectors. The conclusion was that at least for the proposed matrices, the up- and down-sectors have rather similar textures, which is not so surprising, given that the weak mixing matrix $V_{CKM}$, being the ``bridge'' between the two charge sectors, has a structure that is not that far from the 3 $\times$ 3 unit matrix.
Our point of departure was the democratic matrix, corresponding to a situation where the mass eigenvalues are degenerate. 

An ansatz is but an educated guess based on some assumptions, and in our case the assumption is that the fermionic mass matrices have an underlying democratic texture\cite{Fr}\cite{Demo}, like
\begin{equation}\label{m0}
M_0=\frac{T}{3}\begin{pmatrix}
 1  & 1 & 1\\
 1  & 1 & 1\\
 1  & 1 & 1\\
\end{pmatrix}
\end{equation}
where $T$ has dimension mass. This matrix represents a situation where all the particles within a given charge sector initially have the same Yukawa couplings.  
The argument for this assumption is that in the Standard Model, all fermions get their masses from the Yukawa couplings via the Higgs mechanism, and
since the couplings to the gauge bosons of the strong, weak and 
electromagnetic interactions are identical for all the fermions in a given charge sector, it seems like a natural assumption that they should also have identical Yukawa couplings. 
The mass spectrum $(0,0,T)$ of the democratic matrix (\ref{m0}) moreover
reflects the experimental situation with one very heavy and two much lighter fermions.
In the weak basis the democratic matrix $M_0$ is totally flavour symmetric, in the sense that the weak states of a given charge are indistinguishable (``absolute democracy'').

The spectrum $(0,0,T)$ is a good approximation, but we want three non-zero eigenvalues.
One natural first step is therefore to modify the diagonal matrix elements, 
\[
M=\frac{T}{3}\begin{pmatrix}
 \alpha  & 1 & 1\\
 1  & \alpha & 1\\
 1  & 1 & \alpha\\
\end{pmatrix},
\] 
which gives a matrix that indeed has three non-zero mass eigenstates, $\frac{T}{3}(\alpha-1,\alpha-1,\alpha+2)$, but two of the masses are degenerate. In order to get three different mass eigenstates, more modifications are needed, e.g.
\begin{equation}\label{u}
M=\begin{pmatrix}
K  & A & B\\
 A  & K & B\\
 B  & B & K\\
\end{pmatrix},
\end{equation}
where all the matrix elements $A,B, K$ have dimension mass.
\newline
We now have a situation with three different mass eigenstates, corresponding to three families, meaning that we have both mixing and CP-violation, since
mixing is a feature of non-degenerate families.

In order to find physically realistic mass matrices, we must take into account the constraint from CP-violation, so we look for an up-quark mass matrix and a down-quark matrix that agree with the Jarlskog invariant.
The first step is to make sure that the determinant of the commutator of the two matrices must be non-vanishing, keeping in mind that least one of the mass matrices must be complex. 

We look for two simple mass matrices of the kind studied in \cite{Koi}, 
\begin{equation}\label{dm}
M_u\sim\begin{pmatrix}
 K  & A & B\\
 A  & K & B\\
 B  & B & K\\
\end{pmatrix}
\hspace{6mm}
\text{and}
\hspace{6mm}
M_d\sim\begin{pmatrix}
 L  & X & Y\\
 X  & L & Y\\
 Y  & Y & L\\
\end{pmatrix}
\end{equation}
where the matrix elements $K,A,B,L,X,Y$ all have dimension mass. 
One of the matrices must be complex, but we don't want more than three parameters in each matrix, so we let $M_u$ be the real matrix (\ref{u}), and make 
this simple choice for the down-quark matrix,
\[
M_d=\begin{pmatrix}
 L  & Y & Y-iF\\
 Y  & L & Y\\
 Y+iF  & Y & L\\
\end{pmatrix}
\]
These mass matrices, for the up-quark and down-quark sectors
\begin{equation}\label{ud}
M_u=\begin{pmatrix}
 K  & A & B\\
 A  & K & B\\
 B  & B & K\\
\end{pmatrix}
\hspace{6mm}
\text{and}
\hspace{6mm}
M_d=\begin{pmatrix}
 L  & Y & Y-iF\\
 Y  & L & Y\\
 Y+iF  & Y & L\\
\end{pmatrix},
\end{equation}
have six parameters, $K,A,B,L,Y,F$, and a non-vanishing commutator with determinant 
\[
det (M_uM_d-M_dM_u)=2i\hspace{0.5mm}BF^3(A^2-B^2), 
\]

\section{Matrix invariants}
The information content of a matrix is contained in its matrix invariants, 
\begin{enumerate}
  \item $trace(M)=m_1+m_2+m_3$
\item $e_2(M)=m_1m_2+m_1m_3+m_3m_2=\frac{1}{2}[(trace(M))^2-trace(M^2)]$
\item $det(M)=m_1m_2m_3=\frac{1}{6}[trace(M))^3+2trace(M^3)-3trace(M)trace(M^2)]$, 
\end{enumerate}
In the case of
\[
M_u=\begin{pmatrix}
K  & A & B\\
 A  & K & B\\
 B  & B & K\\
\end{pmatrix}
\]
we use $trace(M)$ and $det(M)$, i.e.
\begin{align*}
&trace(M)=3K\\
&det(M)=K^3+2AB^2-2KB^2-KA^2=(K-A)\left[K(A+K)-2B^2\right]\\
\end{align*}
Since the determinant factorizes, we can read off one eigenvalue directly,
\begin{align*}
&m_1=K-A\\
&m_2m_3= [K(A+K)-2B^2],\\
\end{align*}  
and inserting
\begin{align*}
&K=(m_1+m_2+m_3)/3\\
&A=(m_2+m_3-2m_1)/3\\
\end{align*}
we get
\[
B^2=\frac{1}{2}[K(A+K)-m_2m_3],
\]
thus
\begin{align*}
  &K=(m_1+m_2+m_3)/3  \\
  &A=(m_2+m_3-2m_1)/3  \\
  &B=\frac{1}{3}\sqrt{(m_3-2m_2+ m_1 )(2m_3- m_2-m_1 )/2}
\end{align*}
Since we immediately found one of the eigenvalues, $m_1=K-A$, we can express the determinant as the product $m_1m_2m_3=(K-A)\times m_2m_3$, where the term $m_2m_3$ leads to a second degree equation, the solution of which gives $m_2$ and $m_3$, whereby we get the explicit eigenvalues
\begin{align*}\label{eig} 
&m_1=K-A\\ 
&m_2=(2K+A-\sqrt{8B^2+A^2})/2\\
&m_3=(2K+A+\sqrt{8B^2+A^2})/2
\end{align*}
If we instead solve the characteristic equation, we get the compact expression for the eigenvalues in terms of the matrix invariants:
\begin{equation}\label{m1}
m_j^{(u)}=\frac{Tr(M)}{3}+\frac{2}{3}\sqrt{Tr(M)^2-3e_2(M)}\cos\left[\frac{1}{3}\arccos\left(\frac{2Tr(M)^3-9e_2(M)Tr(M)+27det(M)}{2(Tr(M)^2-3e_2(M))^{3/2}}\right)-\frac{2\pi j}{3}\right]
\end{equation}
where $j=0,1,2$.
\newline
To calculate the matrix elements of the mass matrix for the down-sector, 
\[
M_d=\begin{pmatrix}
 L  & Y & Y-iF\\
 Y  & L & Y\\
 Y+iF  & Y & L\\
\end{pmatrix},
\]
we again use matrix invariants, 
\begin{enumerate} 
\item $trace(M_d)=3L$
\item $e_2(M_d)=3L^2-3Y^2-F^2$  
\item $det(M_d)=L^3+2Y^3-L(3Y^2+F^2)$
\end{enumerate}
From relation 2. we see that $3Y^2+F^2=3L^2-e_2(M_d)$, thus
\[
det(M_d)=L^3+2Y^3-L(3L^2-e_2(M_d))\hspace{1mm}\Rightarrow \hspace{1mm}2Y^3=det(M_d)+2L^3-Le_2(M_d)
\]
and
\[
Y=\left[\frac{det(M_d)+2L^3-Le_2(M_d)}{2}\right]^{1/3}
\]

\section{Mass eigenvalues}
The eigenvalues of the up-quarks were easily found: \[
(m_1,m_2,m_3)=(K-A,(2K+A-\sqrt{8B^2+A^2})/2, (2K+A+\sqrt{8B^2+A^2})/2),\] 
but in order to find the eigenvalues of $M_d$, we must solve
\[
det \begin{pmatrix}
 L-\lambda  & Y & Y-iF\\
 Y  & L-\lambda & Y\\
 Y+iF  & Y & L-\lambda\\
\end{pmatrix}=0
\]
That is,
\[
(L-\lambda)^3+2Y^3-(L-\lambda)(3Y^2+F^2)=0
\]
We substitute $\lambda=L+w$, which gives the cubic equation 
\[
w^3-w(3Y^2+F^2)-2Y^3=0
\]
We make the ansatz 
$w=u \cos\theta$, where
\[
u= 2\sqrt{\frac{3Y^2+F^2}{3}} \text{and} \cos(3\theta)=Y^3(\frac{3}{3Y^2+F^2})^{3/2} 
\]
and we get
\[
m_j=L+2\sqrt{\frac{3Y^2+F^2}{3}}\cos\big{[}\frac{1}{3}\arccos[Y^3(\frac{3}{3Y^2+F^2})^{3/2}]-\frac{2\pi\hspace{1mm}j}{3}
\big{]}
\]
where $j=1,2,3$ and $m_j$ are the down-quark masses $m_d,m_s,m_b$.
\newline
Reformulating $Y$ and $F$ in terms of the matrix invariants, we get
\begin{equation}\label{m2}
  m_j^{(d)}=\frac{Tr(M_d)}{3}+\frac{2}{3}\sqrt{Tr(M_d)^2-3e_2(M_d)}\cos\left[\frac{1}{3}\arccos\left(\frac{2Tr(M_d)^3-9e_2(M_d)Tr(M_d)+27det(M_d)}{2(Tr(M_d)^2-3e_2(M_d))^{3/2}}\right)-\frac{2\pi j}{3}\right]
\end{equation}
so despite the seemingly complicated form of 
$M_d$, its eigenvalues ultimately depend only on the three basis-independent invariants, just as they should. We notice the similarity with (\ref{m1}): the matrices $M_d$ and $M_u$ have eigenvalues of exactly the same structural form, both depending only on their three matrix invariants.
So despite the very different structures of the two matrices, one real, the other complex - their eigenvalues are governed by the same formula, depending only on the three basis-independent invariants. 
In this sense, the texture of the matrix in the flavour basis is physically irrelevant; only the invariants matter.

\section{Numerical matrices}
In order to get a picture of the structure of $M_u$, we want to insert numerical quark mass values in $m_j$ (for our purpose, it is not important that there is some uncertainty in the quark masses). Using these quark mass values \cite{jamin},\cite{Jamin2} at $M_Z$:    
\begin{equation}\label{jamin} 
\begin{matrix}
m_u(M_z)=1.24 \hspace{1mm}MeV, & m_c(M_z)= 624 \hspace{1mm}MeV, & m_t(M_z) = 171550 \hspace{1mm}MeV\\
m_d(M_z)=2.69 \hspace{1mm}MeV, & m_s(M_z)= 53.8 \hspace{1mm}MeV, & m_b(M_z) = 2850 \hspace{1mm}MeV \\
\end{matrix} 
\end{equation}
we get these numerical values for the matrix elements in the up-sector
\[
K=57391.75, \hspace{2mm}A=57390.5,\hspace{3mm} B=56923.2,
\] 
and the mass matrix for the up-quarks shows a nearly democratic texture:
\begin{equation}\label{up}
M_u(M_Z)=\begin{pmatrix}
 57391.75  & 57390.5   & 56923.22\\
 57390.5    & 57391.75 & 56923.22\\
 56923.22  & 56923.22 & 57391.75\\
\end{pmatrix}=56923.22 MeV\begin{pmatrix}
 1.00823  & 1.00820   &1 \\
 1.00820    & 1.00823 & 1\\
 1  & 1 & 1.00823\\
\end{pmatrix} 
\end{equation} 
This allows us to numerically calculate the determinant for the commutator:
\[
det (M_uM_d-M_dM_u)=2i\hspace{0.5mm}BF^3(A^2-B^2),
\]
which we insert into $J_{CP}$ to calculate the numerical value of $F$, 
\[
J_{CP}=-i\hspace{1mm} det[M_u,M_d]/2 P_u P_d=BF^3(A^2-B^2)/P_uP_d=3.096\times 10^{-5},
\]
i.e. $F^3=3.096\times 10^{-5}\times P_uP_d/(B(A^2-B^2))$,
which gives $F=42.295 MeV$.

Inserting the numerical values from (\ref{jamin}) into the matrix invariants of $M_d$, we get \[Y=940.4 MeV,\] and we can write the numerical mass matrices as
\[
M_u(M_Z)=
56923.22 MeV\begin{pmatrix}
 1.00823  & 1.00820   &1 \\
 1.00820    & 1.00823 & 1\\
 1  & 1 & 1.00823\\
\end{pmatrix} 
\]
and
\[
M_d(M_Z)=
940.35 MeV\begin{pmatrix}
 1.03  & 1 & 1-i\hspace{1mm}0.045\\
 1  & 1.03 & 1\\
 1+i\hspace{1mm}0.045  & 1 & 1.03\\
\end{pmatrix} 
\]
which both have a democratic texture.

\section{The reduction of parameters}
We can express the up quark matrix elements in terms of the up-quark masses:   
\begin{description} 
\item$K=(m_u+m_c+m_t)/3$ 
\item$A=(m_c+m_t-2m_u)/3$ 
\item$B=\frac{1}{3}\sqrt{(m_t - 2 m_c + m_u) (2 m_t - m_c - m_u)/2}$
\end{description} 
Likewise, the down sector has matrix elements
\begin{description} 
\item$L=(m_d+m_s+m_b)/3$
\item$Y=\left[\frac{det(M_d)+2L^3-Le_2(M_d)}{2}\right]^{1/3}$
\item$F=\left[3.096\times 10^{-5}\times P_uP_d/(B(A^2-B^2))\right]^{1/3}   $  
\end{description} 
We see that the last down-quark matrix element is a function of the up-quark matrix elements $A$ and $B$, which allows us to reformulate $A$:
\[
A^2-B^2=3.096\times 10^{-5}\frac{P_uP_d}{F^3B} \hspace{4mm}\Rightarrow \hspace{4mm}
\]
\[
A=\sqrt{\frac{(BF)^3+3.096\times 10^{-5}P_uP_d}{F^3B}}\]
where $P_u=(m_u-m_c)(m_c-m_t)(m_t-m_u)$ and $P_d=(m_d-m_s)(m_s-m_b)(m_b-m_d)$.
Our two mass matrices are now defined by five parameters, $K,B,L,Y,F$.

So the mass eigenvalues for the up-sector are expressed in terms of $K,B,F$, while the mass eigenvalues for the down-sector are expressed in terms of $L,Y,F$, 
i.e. the mass eigenvalues of the two sectors are not independent of each other, but intertwined.

\section{Conclusion}
We have shown that the constraint imposed by CP-violation, specifically the Jarlskog invariant, links the mass matrices of the up-quarks and down-quarks. For the concrete ansatz (\ref{ud}), this constraint reduces the number of independent matrix parameters from six to five, which means that the mass eigenvalues of the two sectors are not independent but intertwined. This is explicitly demonstrated by expressing the up-sector matrix
$A$, 
\[
A=\sqrt{\frac{(BF)^3+3.096 \times 10^{-5} P_uP_d}{F^3B}}
\]
where 
$P_u=(m_u-m_c)(m_c-m_t)(m_t-m_u)$, $P_d=(m_d-m_s)(m_s-m_b)(m_b-m_d)$, and $B$ and $F$ are matrix elements in the up-sector and down-sector matrices, respectively.
Although this interdependence is demonstrated here through a specific ansatz, the entanglement of the two sectors via the CP-violation constraint is a general feature, independent of the particular model chosen.

\end{document}